\documentstyle[12pt,amsfonts,epsfig]{article}
\begin{document}
\def\k{{k_F}}
\centerline{THE LONGITUDINAL AND TRANSVERSE RESPONSES}
\centerline{IN THE INCLUSIVE ELECTRON SCATTERING}
\vskip1cm
\centerline{\underline{R. Cenni}, F. Conte and P. Saracco}
\vskip0.6cm
\centerline{Istituto Nazionale di Fisica Nucleare}
\centerline{Dipartimento di Fisica -- Universit\`a di Genova}
\centerline{Via Dodecaneso 33 -- 16146 -- Genova -- Italy}
\vskip0.6cm\hskip2cm
ABSTRACT
\newline
{\em The splitting between the charge-longitudinal and spin-transverse 
responses is explained in a model whose inputs are the effective 
interactions in the particle-hole channels in the first 
order boson loop expansion. 
The 
interplay between $\omega$-meson exchange and box diagrams 
mainly governs the longitudinal response, while
in the transverse one the direct $\Delta$ excitations almost cancel the 
one-loop correction and the response is ruled by the 
$\rho$-meson rescattering.}

\section{The experimental and theoretical situations\label{sect2}}

The experimental outcomes in the quasi-elastic peak (QEP) region 
are at present still controversial both on the
experimental and theoretical point of view.

Starting from 
$\frac{d^2\sigma}{d\Omega d\epsilon}=\sigma_M\left\{v_LR_L(q,\omega)+v_T
R_T(q,\omega)\right\}$
the Saclay experimentalists \cite{Me-al-84,Me-al-85} where   
able to perform the 
Rosenbluth separation thus getting both $R_L$ and $R_T$. 
The longitudinal response
was 
drastically quenched with respect to the Free Fermi Gas (FFG) model, 
while the transverse one was remarkably increased.

The first difficulties came from the 
non-fulfillment of the Coulomb sum rule, that, being
expected to provide the nuclear charge, was 
quenched to a, say, 90\% in case of ${}^{12}C$ but to a 60\% in the case of 
$^{40}Ca$.

Few years ago the Rosenbluth separation has also been performed 
at Bates 
\cite{Ya-al-93}. The quenching of the sum rule for the $^{40}Ca$ turned out to 
be of about a 10\%, in sharp contrast with 
the Saclay data. 

Very recently 
Jourdan \cite{Jo-95} 
outlined that the 
Rosenbluth separation is not  free from theoretical ambiguities, and 
others are introduced in deriving the sum rule:
the distortion of the outgoing electron must be 
correctly accounted for, before separating the 
channels; further, relativity prevents us to define the Coulomb sum 
rule in a natural way \cite{Ba-al-94}.
Jourdan showed that the corrected 
sum rule derived from world set of data is compatible with $Z$ within a 
1\% incertitude.

This outcome is, in principle, not strongly contradictory with the Saclay 
results: the Coulomb sum rule in fact properly reads
$
\int\limits_0^\infty{R_L(q,\omega)}/{G_E^2(q^2)}\simeq 
S_L(q)=Z+\frac{2Z}{\rho}g(q)\;,
$
$g(q)$ being the pair correlation function, and the outcome of \cite{Jo-95} 
only states that $g(q)$ is compatible with 0 at, say, $q=$570 MeV/c.
At lower momenta still a 
sizeable quenching 
survives.

Coming to the theory,
all approaches agree in 
providing a more or less pronounced depletion of the QEP in the 
longitudinal channel \cite{FaPa-87,CoQuSmWa-88,Ba-al-95,VaRyWa-95}. 
The transverse response has been less 
investigated. 
Ref. \cite{VaRyWa-95} 
seems to be (up to our knowledge) the only attempt to explain 
simultaneously the two response within the same frame, namely that of 
continuum RPA plus two-body currents. Still, both responses 
seem to be slightly overestimated. 

\section{The theoretical frame \label{sect3}}

\begin{figure}
\begin{center}
\mbox{
\epsfig{file=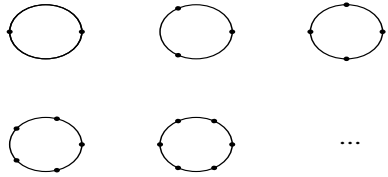,height=3cm,width=6cm}
}
\end{center}
\caption{\protect\label{fig1} The bosonic effective action. Dots denote
the external points.}
\end{figure}

Our starting point in \cite{AlCeMoSa-87} was to build up a 
well-behaved approximation scheme for a system of nucleons and pions.
The idea of the bosonization arose quite naturally there,
as it was obtained by representing the 
generating functional of the system by means of a Feynman path integral 
and by integrating over the fermionic degrees of freedom. 
Then the system turns out to be described by the bosonic 
effective action
\begin{eqnarray}
\lefteqn{S_B^{\rm eff}=\frac{1}{2}\int 
d^4x\,d^4y\,\phi(x)\left[D_0\right]^{-1}(x-y)
\phi(y)}
\label{lt5}\\
&&-\sum_{n=2}^\infty\int d^4x_1\dots d^4x_n
\frac{1}{n}\Pi^{(n)}(x_1,\dots,x_n)\phi(x_1)\dots \phi(x_n)\;,
\nonumber
\end{eqnarray}
$\phi$ denoting the pion field,
$D_0$ its free propagator and the nonlocal vertices 
$\Pi^{(n)}(x_1,\dots,x_n)$,
that keep memory of the fermion dynamics, being shown
in fig. \ref{fig1}. 
\begin{figure}
\begin{center}
\mbox{
\epsfig{file=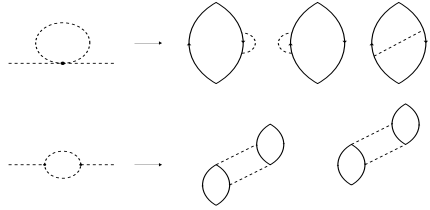,height=3cm,width=6cm}
}
\end{center}
\caption{\protect\label{fig2} First order diagrams in the BLE. Recall that the
dashed lines always describe RPA-dressed bosons}
\end{figure}
\begin{figure}
\begin{center}
\mbox{
\epsfig{file=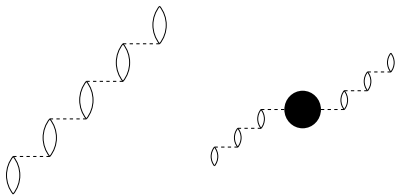,height=3cm,width=6cm}
}
\end{center}
\caption{\protect\label{fig2bis} Diagrams pertaining to the response 
function at the one-loop level. The black bubble summarizes the 5 
diagrams of fig.\protect\ref{fig2}
dashed lines always describe RPA-dressed bosons}
\end{figure}
This action generates a new class of approximations -- or a recipe to 
collect classes of Feynman diagrams together -- when the semiclassical 
expansion is carried out. This scheme is referred to as boson loop 
expansion (BLE). 

The recipe to classify a Feynman diagram according to its order 
in BLE is to shrink to a point all closed fermion lines 
and to count the number of boson loops left out.

Thus the mean field level coincides with the RPA, as
can be seen diagrammatically.
At the linear response level the mean field 
is thus described by the RPA-dressed polarization propagator if the 
probe has the same quantum numbers of the pion or, if not, by the bare Lindhard 
function.
At the one-loop order the only possible diagrams 
are those of fig. \ref{fig2}. 
The full response at the one-loop order is given 
by the imaginary part of the diagrams of fig. 
\ref{fig2bis}, where the black bubble denotes the sum of all the 
diagrams of fig. \ref{fig2}.

It is evident from fig. \ref{fig2} that only fermion loops with at most four 
external legs intervene at the one-boson-loop order. This fact is of 
overwhelming relevance since analytical expressions are available for them 
\cite{CeSa-88,CeCoCoSa-92} at least in the nonrelativistic kinematic.

The pion 
condensation is unavoidably met in the RPA scheme
if we only allow pion exchange without accounting for short 
range correlations (SRC). 
To avoid it 
we are forced to forbid its occurrence 
by phenomenologically embedding SRC in our 
model by means of the
Landau parameter $g'$, that here 
plays the role of a potential.

To adhere to phenomenology we are thus forced to change somehow our 
approach
and to adopt a potential frame: 
$D_0^{-1}$ is replaced by the inverse 
potential in the given p-h channel (meson exchange plus Landau 
parameter) and the field $\phi$ is
reinterpreted as an auxiliary field. The topology of the diagrams 
remains unchanged.

In going from the mean field to the one loop level, a 
qualitative change is met. In the former case 
not too high momenta
enter the dynamics, while in the latter
the loop momentum is integrated over and the high momentum behaviour 
of the effective interaction rules the convergence of the integral.
This is exploited 
by forcing a $q$-dependence in 
$g^\prime$ such that 
$g^\prime(q)\stackrel{q\to\infty}{\longrightarrow}1\;.$
But this in 
turn requires the introduction of a further cut-off related to the SRC 
range and which becomes the crucial parameter ruling the one loop 
corrections. 

So far we have only  discussed a system of nucleons and only one 
effective interaction. The dynamics required to describe
the nuclear responses is by far richer. 
Other channels are
accounted for by simply interpreting the dashed lines in figs. \ref{fig2} 
as a sum over all the allowed channels.
Also, the excitation of the nucleons to a $\Delta$ 
can be 
allowed by interpreting each solid line as a nucleon or as a $\Delta$ 
and again summing over all possible cases. {\em Remarkably the topology 
of the diagrams does never changes}. 

\section{The dynamical model \label{sect4}}

We shall consider here the (correlated) exchange of $\pi$, 
$\rho$ and $\omega$-mesons.
In each particle-hole channel the interaction will read
\begin{equation}
V_m(q)={f^2_{\pi BB^\prime}\over m_\pi^2}\left\{g^\prime_m({\bf q})-
C_m{{\bf q}^2\over m^2_m+{\bf q}^2}\right\}v_m^2(q^2)\;,
\label{lt3}
\end{equation}
where the index $m$ runs over the accounted channels and $B=N$ or 
$\Delta$.

Concerning the pion, 
by definition, we have $C_\pi=1$. 

A vector meson interact with the nucleons 
via a Coulomb-like plus a spin-transverse interaction plus eventually an 
anomalous spin current.
Customarily the convective 
current is neglected. 

Since the 
``anomalous magnetic moment'' is absent in the $\omega$ case, the 
coupling constants of the Coulomb-like and spin-transverse interactions 
would coincide, while
the anomalous spin current of the $\rho$ will dominate.
We shall also account for the 
interaction of the $\rho$ meson with the nuclear medium other than ph 
and $\Delta$h excitations (we means for instance $\rho\to\pi\pi$ with 
further interaction of pions with the medium) by attributing to the 
$\rho$ (in the spin-transverse channel only) an effective mass.

Remarkably the $\rho$-exchange, due to its high mass and its vector 
character, is essentially perturbative and the coupling constant can be 
invoked from the phenomenology, while
coming to the $\omega$ meson, we have left the interaction in 
the spin-transverse channel unchanged, since it is 
weak, but the Coulomb-like interaction needs to be drastically renormalized 
in the medium.

The Landau parameter $g^\prime$ is present in the isovector
spin-transverse and spin longitudinal channels only, where
it has been parametrized according to
\begin{equation}
g^\prime_{L,T}(q)=C_m+(g_0^\prime-C_m)\left[{q_{c\,{L,T}}^2\over
q_{c\,{L,T}}^2+q^2}\right]^2
\label{lt8}
\end{equation}
in such a way that for $q\to 0$ $g^\prime_{L,T}(q)\to g^\prime_0$ and for 
$q\to\infty$ the right behaviour is achieved. The further 
cut-offs are called here $q_{c\,{L,T}}$. 

\section{The mean field level \label{sect5}}

The mean field level in a coherent discussion is expected to precede the 
first order corrections, and we shall follow this aptitude in the present 
exposition. The situation at hand however is particularly unlucky, 
because the mean field level is inextricably linked in the present case 
to the one-loop corrections. In fact:
\begin{enumerate}
\item
In the charge-longitudinal channel the mean field is mainly dominated by 
the $\omega$ exchange, which is repulsive and so entails a quenching of 
the response. As a simple mathematics shows this implies also a 
hardening of the response \cite{AlCeMo-89} that is not observed in the 
data. Further, we neglected in our approach the $\sigma$-meson exchange, 
which is known from the Dirac phenomenology \cite{Wa-74,ClHaMe-82,Cl-86}
to be attractive and that cancels to a large amount with the $\omega$ 
exchange. In our approach, in the same line of thought of the Bonn 
potential \cite{MaHoEl-87,Ho-81} the $\sigma$-meson is replaced 
by the box diagrams, i.e., two meson exchange with simultaneous 
excitation of one or two nucleons to a $\Delta$ resonance. These 
diagrams are contained indeed in our correlation diagrams,
which are however at the one-loop level.
\item
In the transverse channel the one-loop corrections are strongly 
suppressed (we shall see later that the $\Delta$-excitation makes this 
job). Thus the mean field becomes dominant in explaining the response
and in turn it is ruled by the effective interaction in the $\rho$ 
channel. Thus in view of our assumption (suppression of one-loop 
corrections) we can extract information about
the effective interaction from the data.
\end{enumerate}

\section{The one-loop corrections \label{sect6}}

We finally come to the one-loop corrections.

First, let us define once forever the sets of parameters we shall use.

\begin{figure}
\begin{center}
\mbox{
\begin{tabular}{ccc}
\epsfig{file=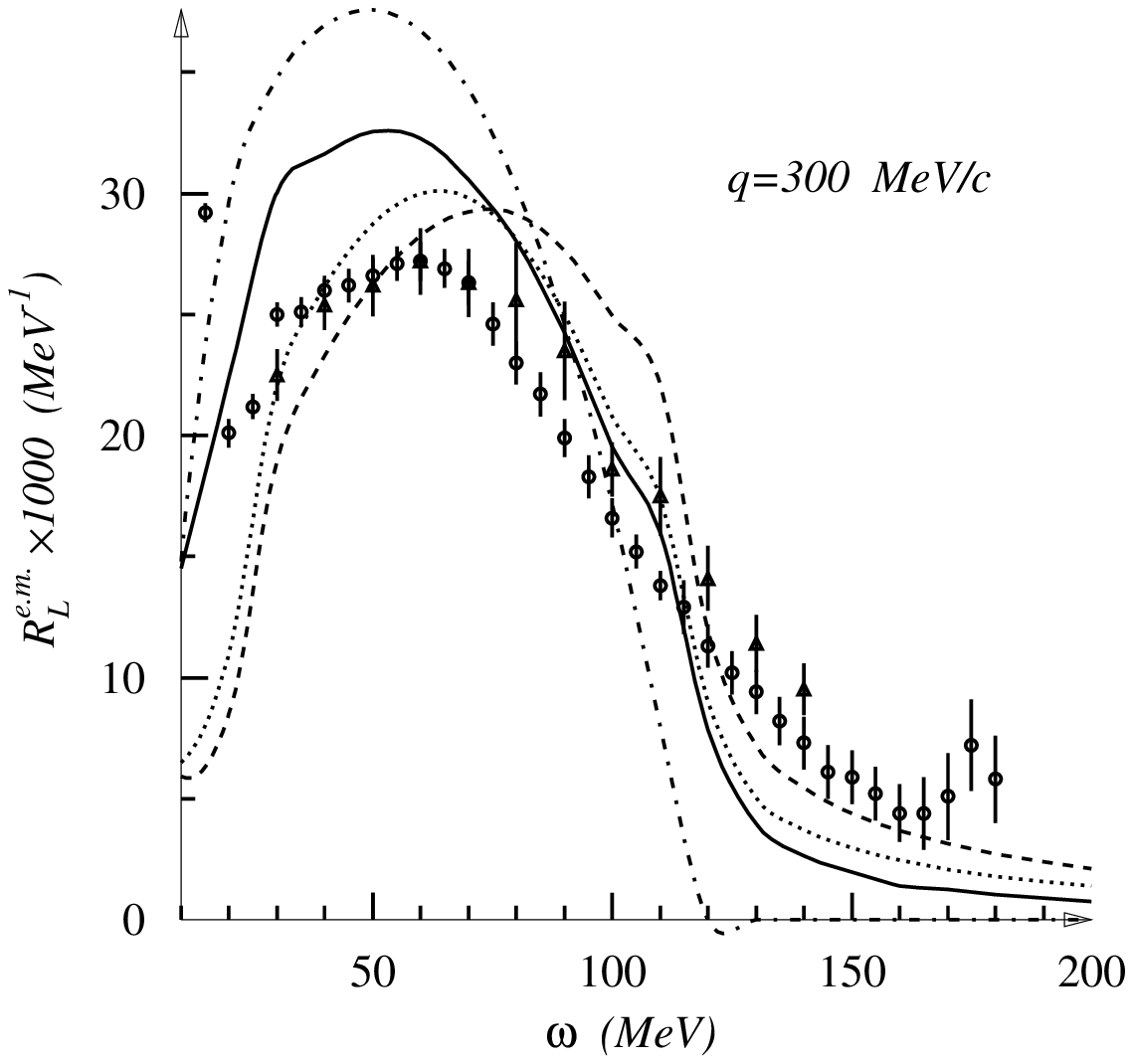,height=6cm,width=4.5cm}&
\epsfig{file=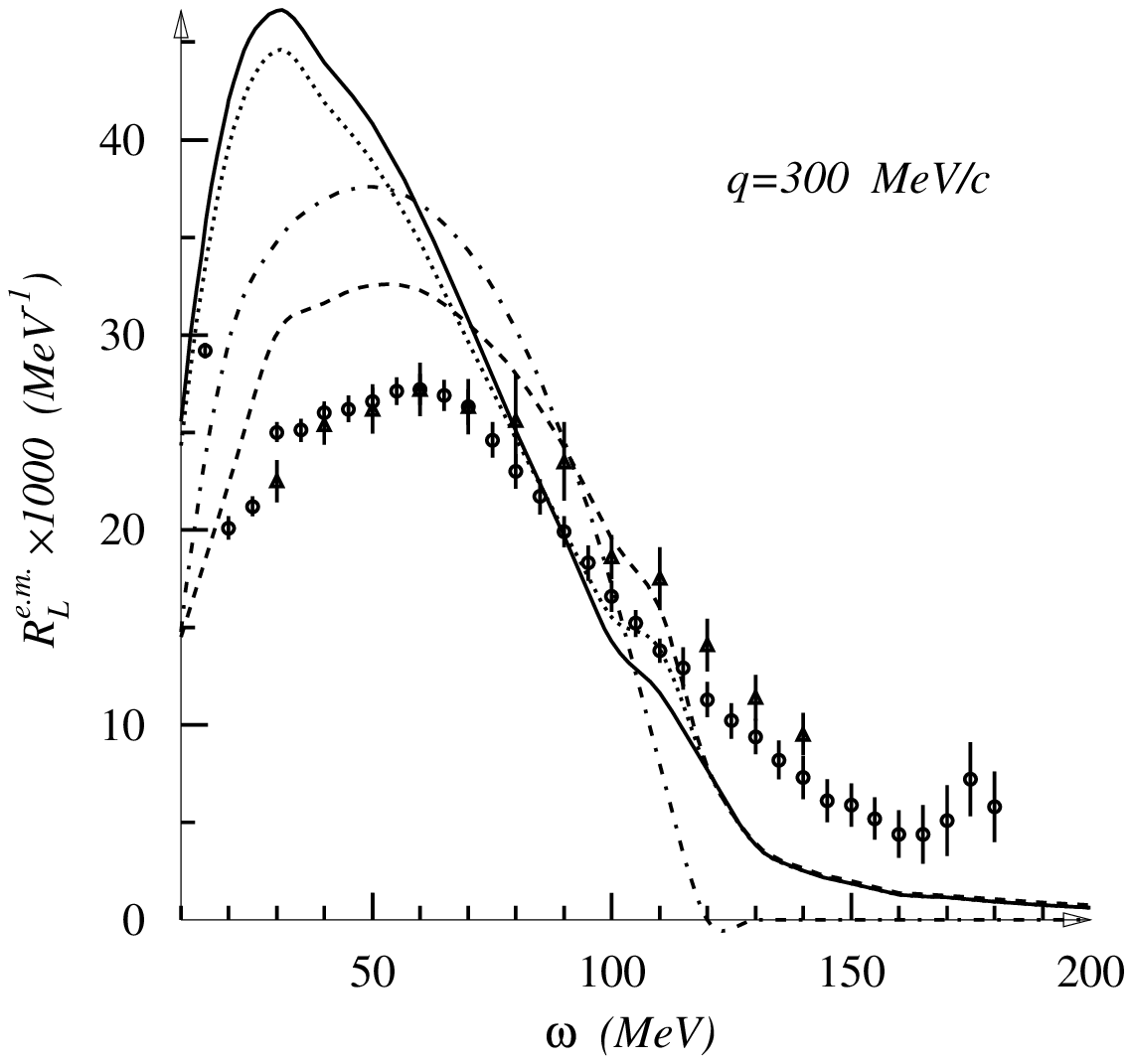,height=6cm,width=4.5cm}&
\epsfig{file=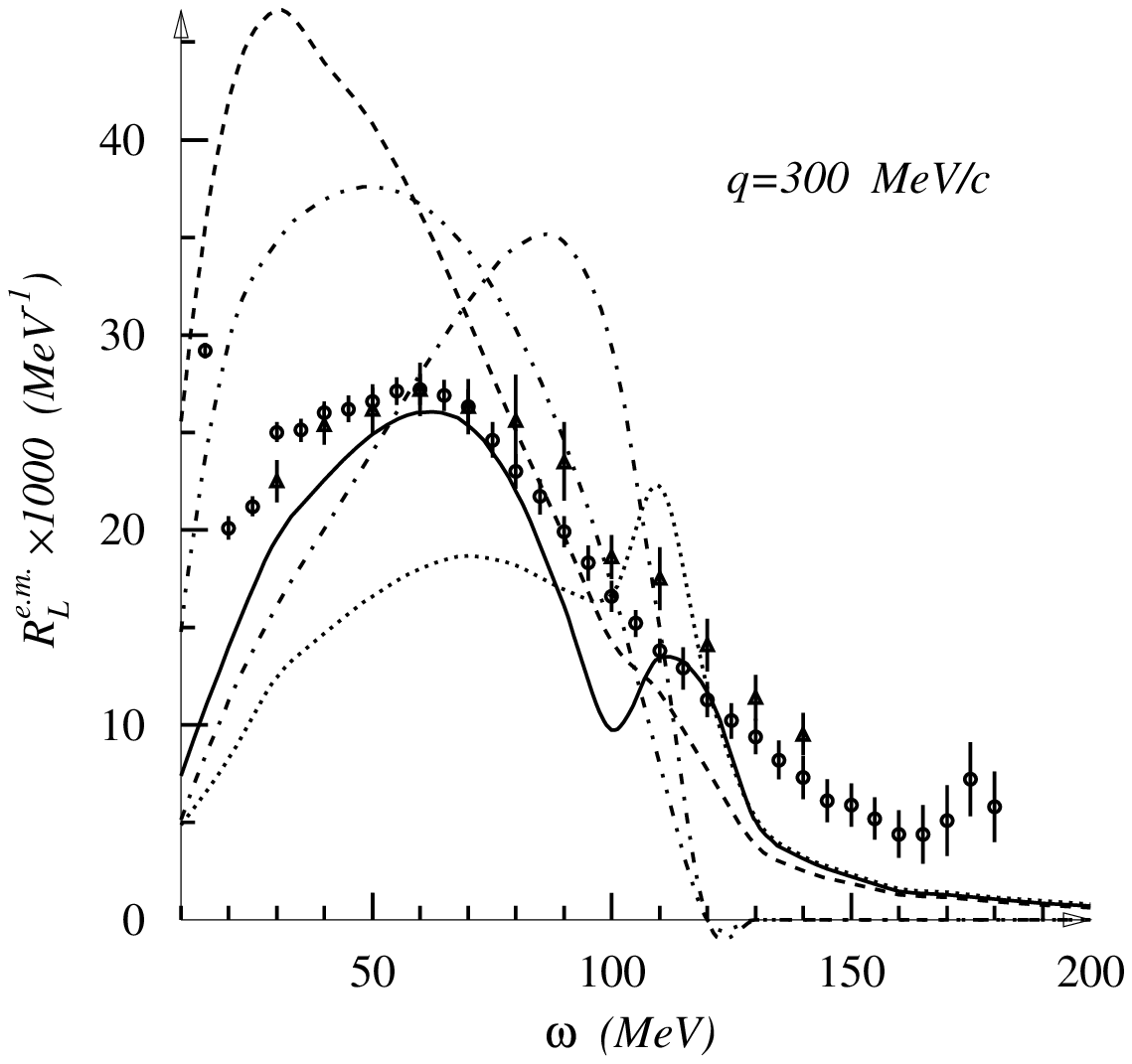,height=6cm,width=4.5cm}
\end{tabular}
}
\end{center}
\caption{\protect\label{fig11}Left: the 
longitudinal response for $^{12}C$ without RPA dressing of the mean 
field and without $\Delta$'s. Dash-dotted line: FFG; dashed line: FFG plus 
self-energy diagrams; dotted line: FFG plus self-energy and exchange 
diagrams; solid line: FFG plus self-energy, exchange and correlation 
diagrams.
\protect\newline
Center: as before but with $\Delta$'s.
Dash-dotted line: FFG; dashed line: diagrams without $\Delta$;
dotted line: box diagrams added
diagrams; solid line: full calculation.
\protect\newline
Right: the full response.
Solid line: full response; dashed line: all diagrams without RPA;
dotted line: diagrams without $\Delta$'s only but with RPA; dash-dotted 
line: mean field; dash-dotted-dotted line: 
FFG}
\end{figure}
The mass of the $\rho$-meson has been set to 600 $MeV$ in the 
spin-transverse channel.
The pion coupling constant have been assumed, as usual, as
$\frac{f^2_{\pi NN}}{4\pi}=0.08$, $\frac{f^2_{\pi N\Delta}}{4\pi}=0.32$ 
and $\frac{f^2_{\pi \Delta\Delta}}{4\pi}=0.016$. 
Next, $C_{\rho NN}=C_{\rho N\Delta}=C_{\rho\Delta\Delta}=2.3$
in the spin-transverse channel and $=0.05$ in the scalar-isovector one.
Further, $C_{\omega NN}=C_{\omega \Delta\Delta}=0.15$ in the 
scalar-isoscalar channel and $=1.5$ in the isoscalar spin-transverse one
(all these results are in agreement with those of the Bonn potential, 
except $C_{\omega NN}$ in the
scalar-isoscalar channel, that is essentially a free parameter).
$g_0^\prime$ is set to 0.35.
The many-body cut-off of SRC are put to $q_{c,L}=800 Mev/c$ and
$q_{c,T}=1300 Mev/c$.
The pion cut-off at the vertices are set to
$\Lambda_{\pi NN}=1300 Mev/c$, $\Lambda_{\pi 
N\Delta}=\Lambda_{\pi\Delta\Delta}=1000 Mev/c$.
The $\rho NN$ cut-off at the vertex in the 
spin-transverse channel is $\Lambda_{\rho NN}=1750 
Mev/c$ and finally
all the remaining cut-offs at the vertices are set to $1000 
Mev/c$.

Let us  start with the longitudinal response and examine, as a guideline, 
the case of Carbon at a transferred momentum of $300~ MeV/c$.

First let us examine the case of pure nucleon dynamics (no $\Delta$'s) 
and also drop RPA dressing
(this means that only the black bubble of fig. \ref{fig2bis}, 
corresponding to the five diagrams of fig. \ref{fig2} survives). 
The main 
effects come from the isovector spin-longitudinal and spin-transverse 
channels, that turn out to be rather similar, while the other channels 
give a negligible contribution. In fig. \ref{fig11}(left) these one-loop 
corrections are shown, and, while going in the right direction, 
are still not sufficient to explain the quenching of the peak. 

\begin{figure}
\begin{center}
\mbox{
\begin{tabular}{ccc}
\epsfig{file=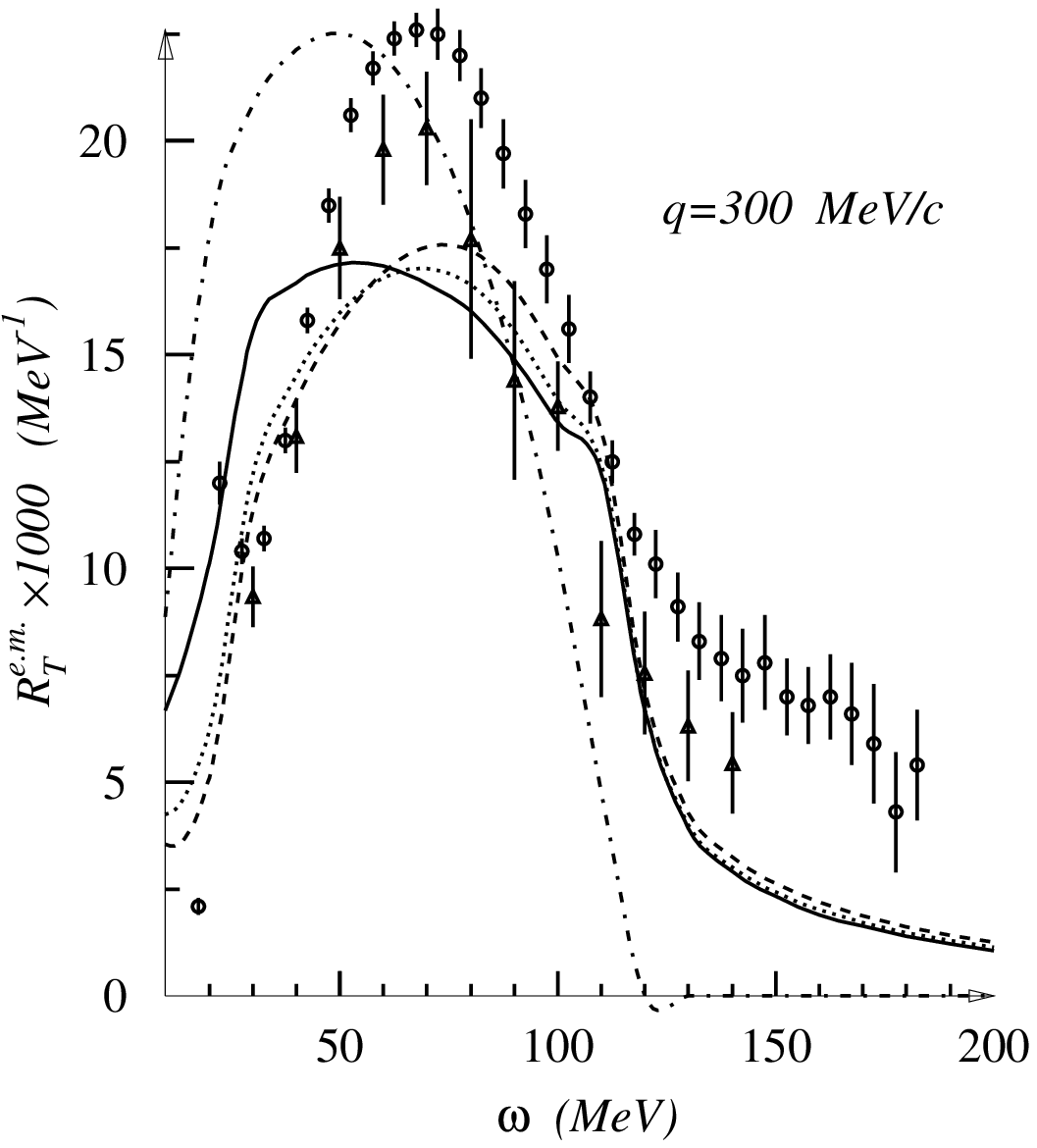,height=6cm,width=4.5cm}&
\epsfig{file=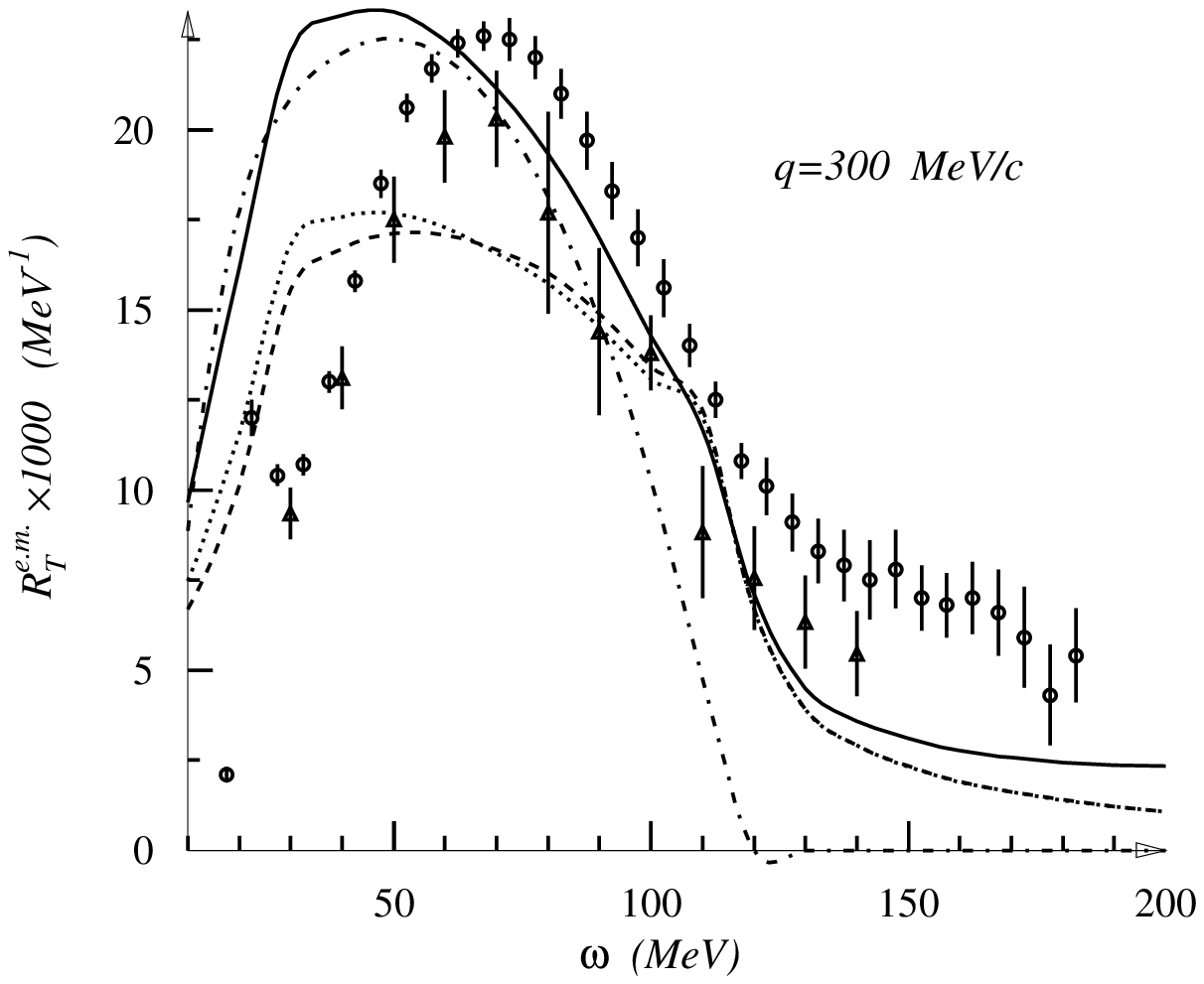,height=6cm,width=4.5cm}&
\epsfig{file=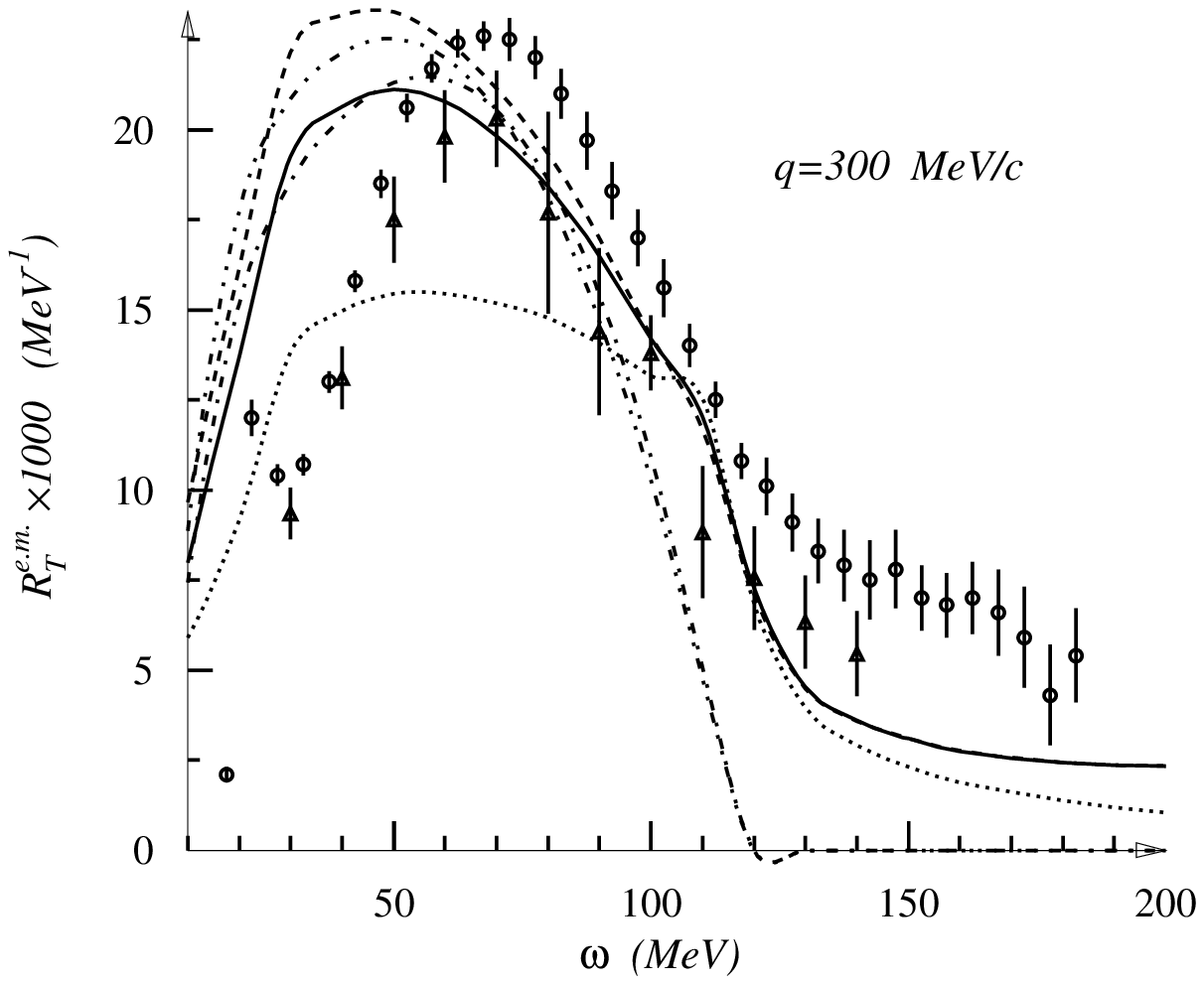,height=6cm,width=4.5cm}
\end{tabular}
}
\end{center}
\caption{\protect\label{fig15}
Left: the 
spin-transverse response for $^{12}C$ without RPA dressing of the mean 
field and without $\Delta$'s. Dash-dotted line: FFG; dashed line: FFG plus 
self-energy diagrams; dotted line: FFG plus self-energy and exchange 
diagrams; solid line: FFG plus self-energy, exchange and correlation 
diagrams.
\protect\newline
Center: as before but with $\Delta$'s.
Dash-dotted line: FFG; dashed line: diagrams without $\Delta$;
dotted line: box diagrams added
diagrams; solid line: full calculation.
\protect\newline
Right: the full response.
Solid line: full response; dashed line: all diagrams without RPA;
dotted line: diagrams without $\Delta$'s only but with RPA; dash-dotted 
line: mean field; dash-dotted-dotted line: 
FFG}
\end{figure}

The next step is to introduce the $\Delta$-resonance. On physical 
grounds we expect that box diagrams will dominate the response. 
They are shown in fig. \ref{fig11}(center) 
together with a complete calculation, 
i.e., with all possible diagrams. The box diagrams are 
dominant indeed and that they make the job the $\sigma$-meson is 
expected to do, namely to strongly enhance and (more important) to 
soften the peak. 
Finally in fig. \ref{fig11}(right) 
we plot the complete graph, with the inclusion 
of the RPA dressing everywhere. 
The net effect is clearly to cancel the enhancement induced by the box 
diagrams in such a way to come back to the experimental data, thus 
accomplishing the expected compensation between $\sigma$- and 
$\omega$-meson as expected from Dirac phenomenology.

Next we examine the transverse response, following the same path as 
before. 
In fig. \ref{fig15}(left) the one-loop corrections are drawn, without 
RPA dressing of the mean field and without $\Delta$'s. The path is 
clearly similar to that fig. \ref{fig11} and is even more 
pronounced, but in the wrong direction, however. 

The contribution of the $\Delta$ diagrams is 
then shown in fig. \ref{fig15}(center), again without RPA-dressing. This time an 
important difference arises, concerning the contribution of the box 
diagrams, that are almost irrelevant. 
Finally in fig. \ref{fig15}(rught) the whole calculation is reported.

Here the physical insight are clear: since one-loop corrections are 
negligible, we are now sensitive to the RPA-dressing of the $\rho$-meson 
channel (the dominant one): we require here an effective interaction 
near to vanish, in order not to further deplete the peak. The one we 
have chosen is still a little bit repulsive: clearly the momentum region 
between 300 and 400 MeV/c is the most sensitive to the details of the 
effective interaction. As a matter of fact our results stay more or less 
between Saclay's and Jourdan's data.

\begin{figure}
\begin{center}
\mbox{
\begin{tabular}{cc}
\epsfig{file=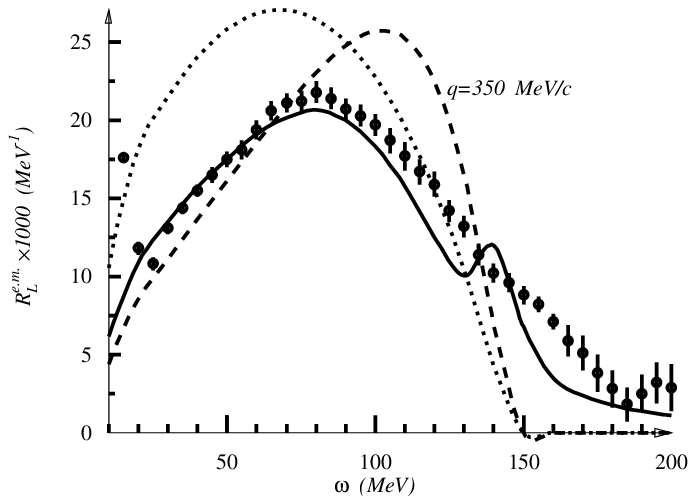,height=4.5cm,width=5cm}
&
\epsfig{file=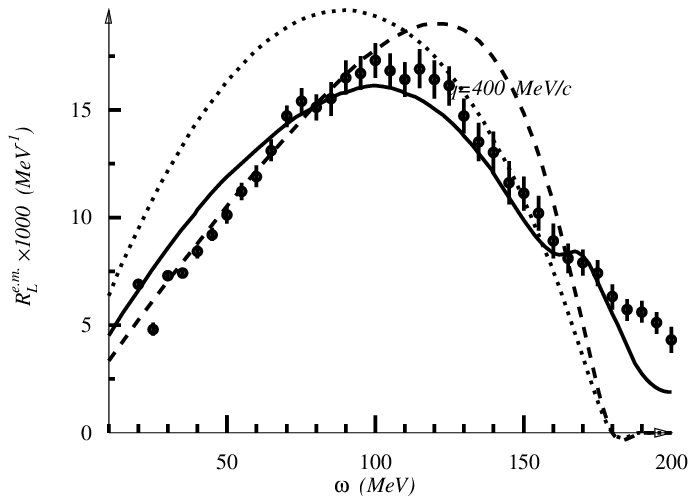,height=4.5cm,width=5cm}
\cr
\epsfig{file=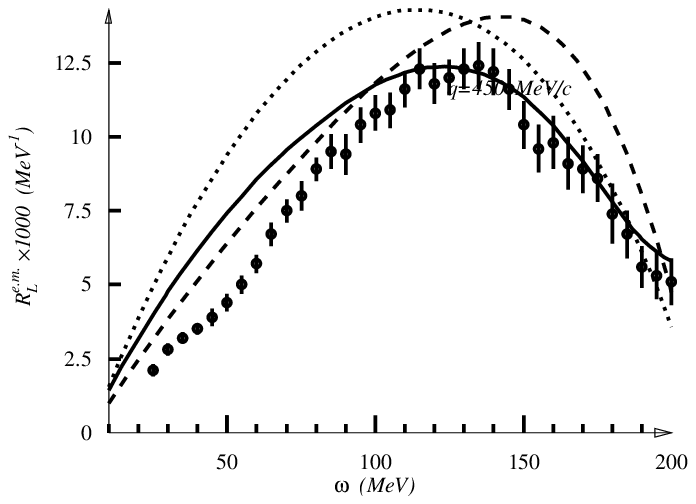,height=4.5cm,width=5cm}
&
\epsfig{file=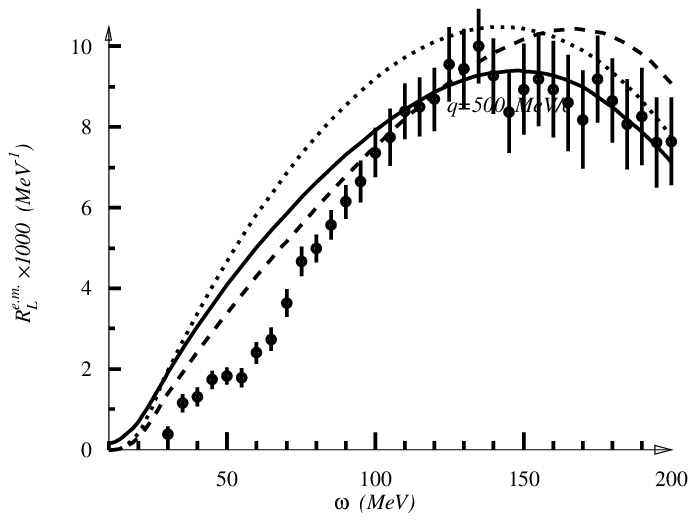,height=4.5cm,width=5cm}
\cr
\epsfig{file=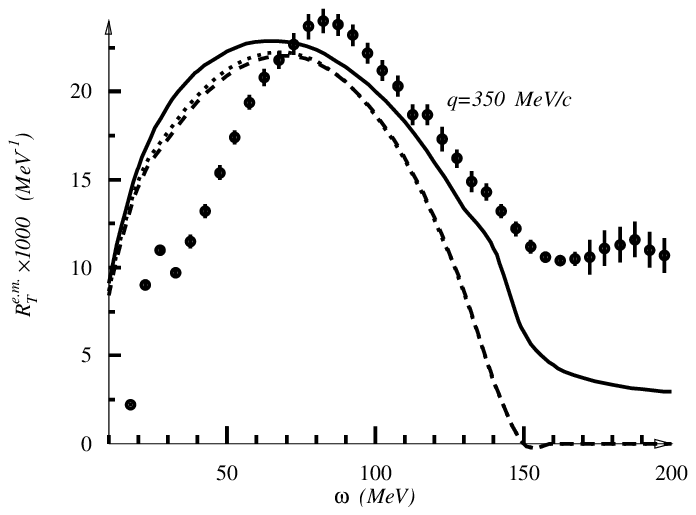,height=4.5cm,width=5cm}
&
\epsfig{file=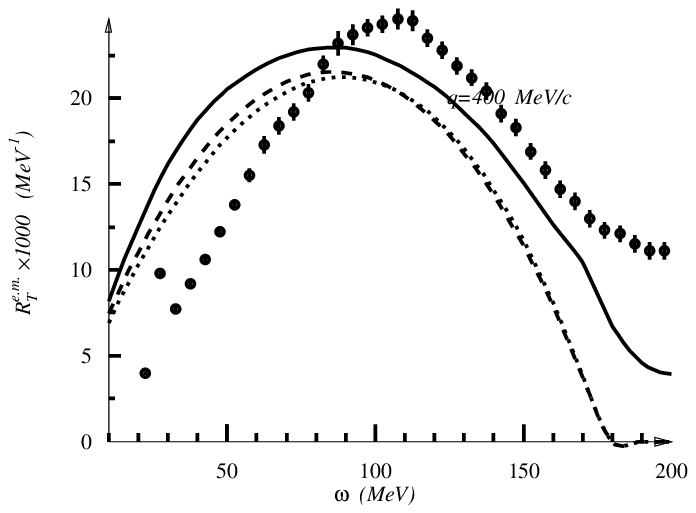,height=4.5cm,width=5cm}
\cr
\epsfig{file=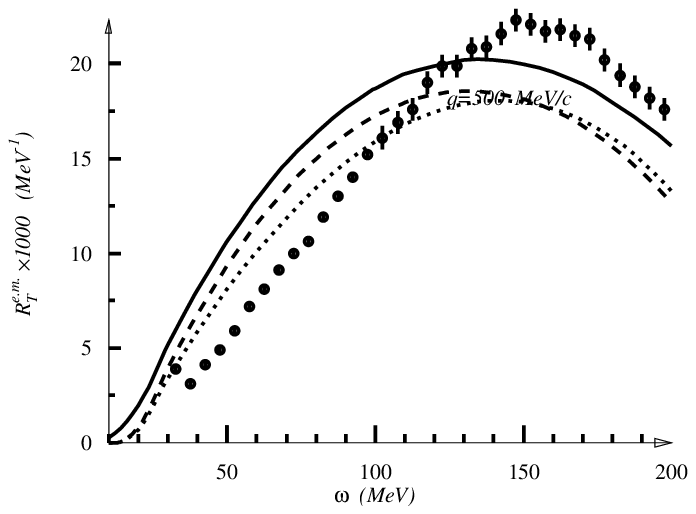,height=4.5cm,width=5cm}
&
\epsfig{file=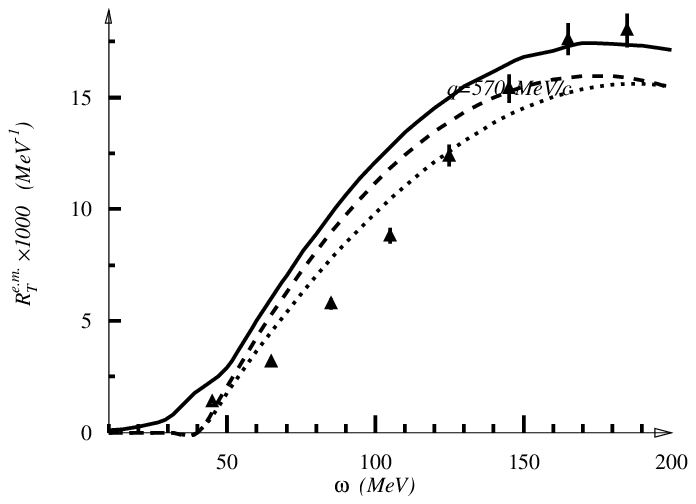,height=4.5cm,width=5cm}
\end{tabular}
}
\end{center}
\caption{\protect\label{fig18} Full calculation for 
the longitudinal (two upper lines) and transverse (two lower lines)
responses on $^{12}C$ at different transferred momenta.
Data from \protect\cite{Me-al-84,Me-al-85} (circles) and from 
\protect\cite{Jo-95,Jo-96} (triangles). Solid line: full calculation
dashed line: Mean field 
calculation
dotted line: FFG calculation.
}
\end{figure}

A more complete survey of our results for $^{12}C$ is presented in fig. 
\ref{fig18}.

\begin{figure}
\begin{center}
\mbox{
\begin{tabular}{cc}
\epsfig{file=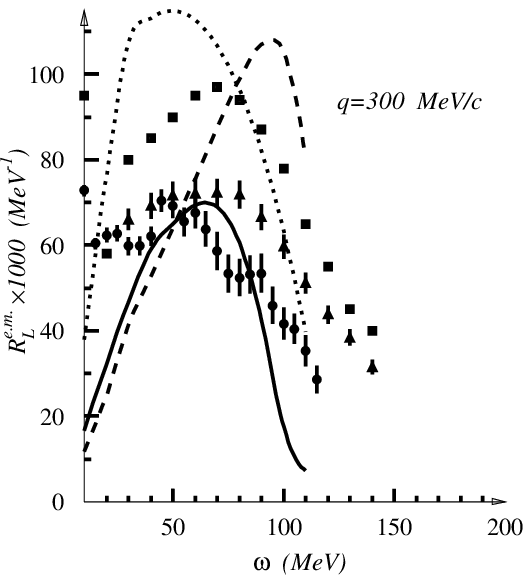,height=4.5cm,width=5cm}
&
\epsfig{file=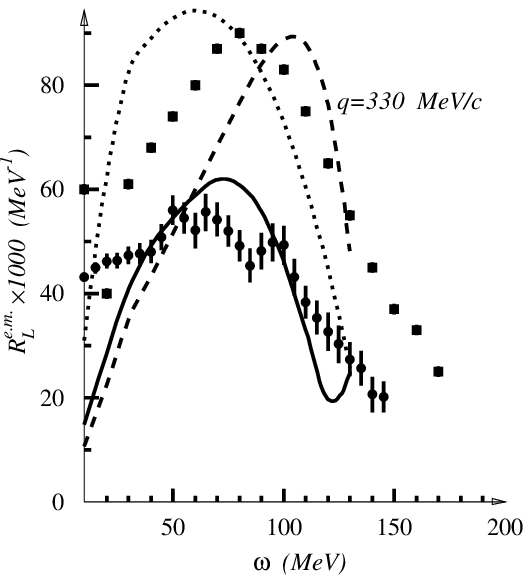,height=4.5cm,width=5cm}
\cr
\epsfig{file=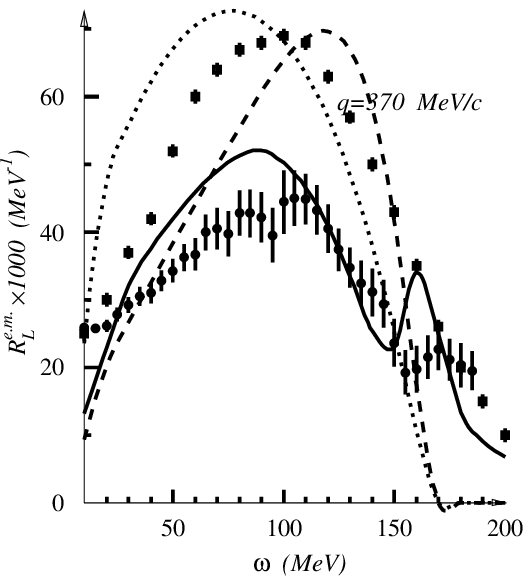,height=4.5cm,width=5cm}
&
\epsfig{file=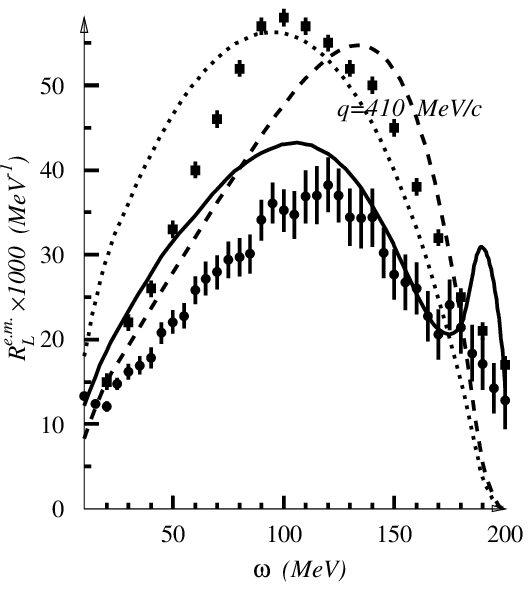,height=4.5cm,width=5cm}
\cr
\epsfig{file=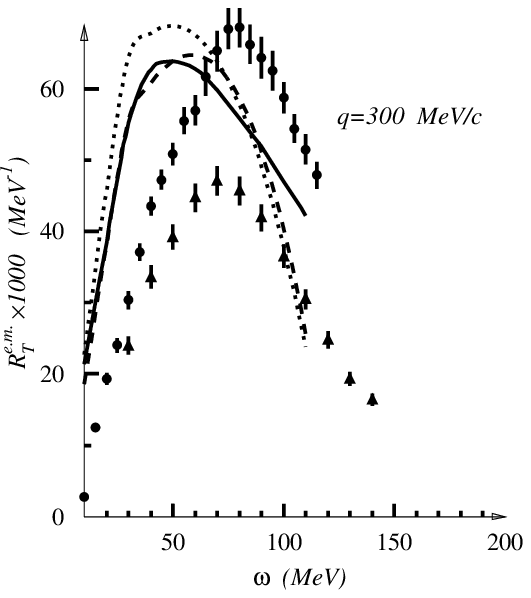,height=4.5cm,width=5cm}
&
\epsfig{file=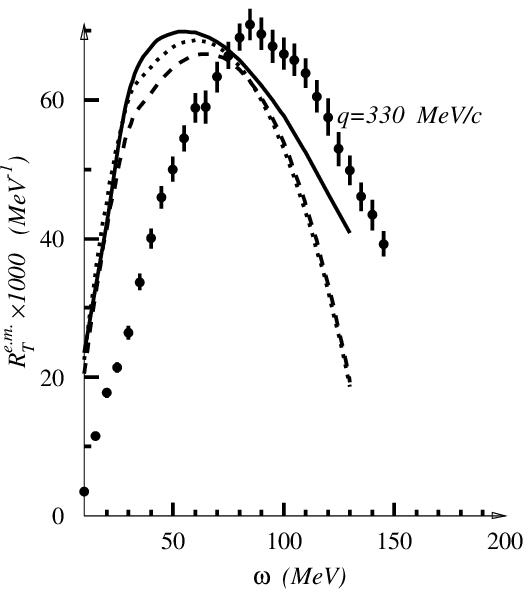,height=4.5cm,width=5cm}
\cr
\epsfig{file=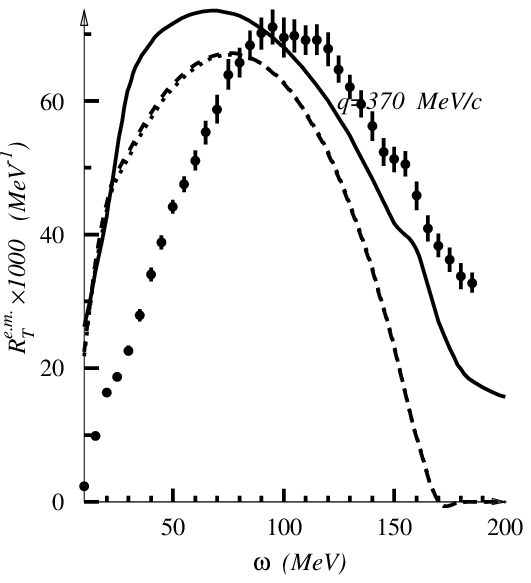,height=4.5cm,width=5cm}
&
\epsfig{file=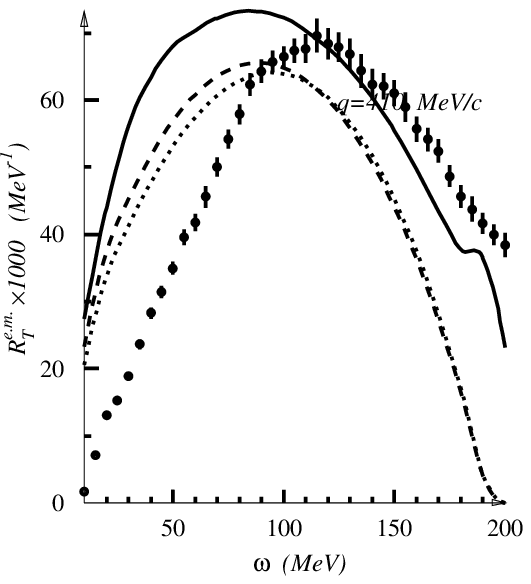,height=4.5cm,width=5cm}
\end{tabular}
}
\end{center}
\caption{\protect\label{fig20} Full calculation for 
the longitudinal (two upper lines) and transverse (two lower lines)
responses on $^{40}Ca$ at different transferred momenta.
Data from \protect\cite{Me-al-84,Me-al-85} (circles) and from 
\protect\cite{Jo-95,Jo-96} (triangles). Solid line: full calculation
dashed line: Mean field 
calculation
dotted line: FFG calculation.
}
\end{figure}

The figures \ref{fig20} present instead the results of 
our calculation ons $^{40}Ca$. In this case we have assumed an higher 
value for $\k$, namely $\k$=1.2~fm$^{-1}$, which better describes a 
medium nucleus. The interplay between RPA dressing of the external legs 
of the diagrams and the one-boson-loop corrections, which also alter the 
real part of the diagram (a density-dependent effect) leads to a more 
pronounced depletion of the longitudinal response, in agreement with 
the Saclay data. This results is relevant in our opinion and deserves a 
comment. 
For a long time the apparent discrepancy between the Saclay 
data on Carbon and Calcium, that provide so different Coulomb sum rule, 
led some people to question about the Calcium data.
We have shown here that such a discrepancy may be explained as a density 
effect in the frame of a well defined theoretical model (the boson loop 
expansion) without violating the Coulomb sum rule. 
\begin{figure}
\begin{center}
\mbox{
\begin{tabular}{cc}

\end{tabular}
}
\end{center}
\caption{\protect\label{fig21} Full calculation for 
the transverse response on $^{40}Ca$ at different transferred momenta.
Data from \protect\cite{Me-al-84,Me-al-85} (circles) and from 
\protect\cite{Jo-95,Jo-96} (triangles). Solid line: full calculation
dashed line: Mean field 
calculation
dotted line: FFG calculation.
}
\end{figure}

Finally, we want to discuss the discrepancies between Saclay and Jourdan 
data. These are particularly emphasized in the transverse response. 
Recalling our previous discussion, the Jourdan data seem to require a 
less pronounced attraction in the isovector spin-transverse effective 
interaction. Since we decreased in our calculations the $\rho$-mass just
to emphasize the attractive part of the interaction 
it is sufficient to keep for the $\rho$-mass its value in 
the vacuum to
better agree with the Jourdan data.
Considering however that still the uncertainty in the experimental 
situation survives and that no microscopical calculations are presently 
available for the $\rho$-meson mass in the nuclear medium, further 
discussions on this topic are still premature.

%\bibliography{[cenni.bibl]references,[cenni.bibl]temp_ref}

\end{document}